\documentclass[twocolumn,caps,showpacs]{revtex4}

\usepackage{graphicx,color}
\usepackage{latexsym}
\usepackage{amsmath,amssymb}        
\usepackage[draft=false]{hyperref}
\usepackage{mathrsfs}
\usepackage{comment}
\usepackage{soul}
\usepackage{mathrsfs}
\DeclareSymbolFontAlphabet{\mathrsfs}{rsfs}
\DeclareMathAlphabet{\mathcal}{OMS}{cmsy}{m}{n}

\begin{document}


\title{Classifying initial conditions of long GRBs modeled with relativistic radiation hydrodynamics}


\author{F. J. Rivera-Paleo, C. E. L\'opez N\'u\~nez, F. S. Guzm\'an, J. A Gonz\'alez}
\affiliation{Laboratorio de Inteligencia Artificial y Superc\'omputo,
	      Instituto de F\'{\i}sica y Matem\'{a}ticas, Universidad
              Michoacana de San Nicol\'as de Hidalgo. Edificio C-3, Cd.
              Universitaria, 58040 Morelia, Michoac\'{a}n,
              M\'{e}xico.}


\date{\today}


\begin{abstract}
We present a method to classify initial conditions of a long gamma ray bursts model sourced by a single relativistic shock. It is based on the use of artificial neural networks (ANNs) that are trained with light curves (LC) generated with radiation relativistic hydrodynamics simulations. The model we use consists in a single shock with a highly relativistic injected beam into a stratified surrounding medium with profile $1/r^2$. In the process we only consider the bremsstrahlung radiation and Thomson scattering process. The initial conditions we use to train the ANN are three: the rest mass density,  Lorentz factor and radiation energy density of the beam that produces the relativistic shock, together with the LC generated during the process. The classification selects the location of a box in the 3d parameter space that better fits a given LC, and in order to decrease the uncertainty of the parameters this box is refined and the classification selects a new box of smaller size.
\end{abstract}


\pacs{07.05.Tp,07.05.Mh,05.45.Tp}


\maketitle

\section{Introduction}
\label{sec:introduction}

There is evidence that long duration gamma ray bursts (GRBs) are produced when massive stars collapse, specifically supernovae with mass bigger than $10 M_\odot$   \cite{piran,woosley,fox,gehrels,kumar}. This evidence comes from two different observations: (1) GRBs have been associated spectroscopically with supernovae of type Ic \cite{woosley}; (2) GRBs are found to be in star-forming regions \cite{fruchter,castro,paczynski}. The connection between observational evidence and theory is important to explain data and validate models. Related to long GRBs, there is a wide variety of models proposed that consider any number of ingredients and processes taking place during a relativistic high energy process \cite{kumar}.

Modeling GRBs, like any other astrophysical scenario involves nowadays a number of processes based on --each time-- more complicated sets of equations, with a growing number of parameters that enrich the models and the scenarios that originated the observations. So far,  models have been constructed in order to infer properties of the GRB source and its surrounding medium. These models are in general numerically solved under the approximation of ideal hydrodynamics, magnetohydrodynamics or radiation hydrodynamics \cite{zhang,mizuta,morsony,bromberg,lazzati,lopez,enrico,Panchos}, and in general define an inverse problem.

Within a model of GRBs there are two important inverse problems. First, the problem of the model, consisting in that given an observed flux light curve (LC) one has to determine the model and processes responsible for the observational data. Second, the problem of the cause, which consists in that given an observed LC, and given a model working under some assumptions (a velocity regime, a density regime, progenitor mass estimates, etc.) one is expected to find the conditions that cause the observational data.

In this paper we focus on the second problem. The model could be as complete as to consider various dispersion processes and elaborate progenitor scenarios \cite{enrico}, or even a full 3D model that works in various frequencies associated to diverse processes happening during the different phases of the evolution of a jet that is candidate for a GRB source \cite{valencia}.

In our case, we assume an idealized model in which the observed LC of a long GRB is due to a single shock propagating at relativistic speed, with its gas coupled to the radiation field \cite{Panchos}. This simple model solves the one dimensional relativistic radiation hydrodynamics equations and the source of the emission is characterized by initial conditions similar to those used in simulations of relativistic jets. Moreover, in our idealized scenario we only consider the opacity associated to bremsstrahlung radiation with Thomson scattering.

The important problem once we have set a model is to find the initial conditions that gave rise to the burst, because they contain the information about the source,  the hydrodynamical processes and the optical conditions of the material of the medium where the jet propagates.

The traditional way in which relativistic shocks and jets are set initially, consider the definition of the beam parameters and those of the surrounding medium. In our case, since the model is one-dimensional we set the jet to be a relativistic shock with initial conditions similar to those of a shock tube, but involving not only the three hydrodynamical variables (density, velocity and energy or pressure), but also the energy and flux of the radiation involved. This means that there are ten parameters defining the initial conditions, five for the beam and five for the surroundings.

Even though the model is one-dimensional and apparently so simple, it has to deal with the numerical solution of a nontrivial set of PDEs for each set of initial conditions. Each solution represents a costly simulation and in the ideal case, if one wants to contrast a model against an observational LC of a given long GRB, the number of simulations required to explore the parameter space should be reduced to the minimum. This is where classification methods are helpful. In this paper we present the use of artificial neural networks (ANN) \cite{Russell,LeCun,backpropagation}, applied to the classification of the initial conditions \cite{carrillo,Raya} of our model of LGRBs based on the one-dimensional relativistic shock with radiation described above.

The paper is organized as follows. In Sec. \ref{sec:methods} we present the numerical methods used to simulate the evolution of the jet producing a GRB according to our model, and the methods related to the ANN. In Sec. \ref{sec:results} we present the results of classification with numerically generated and observational LCs. Finally in Sec. \ref{sec:final} we present our conclusions.


\section{Numerical methods}
\label{sec:methods}


\subsection{Description of the physical model}
\label{subsec:description}

The evolution of a relativistic shock in our scenario, is assumed to be ruled by the radiation relativistic hydrodynamics system, which is a set of PDEs for the gas density $\rho$, the hydrodynamical pressure $P$, the gas velocity $v$ or equivalently its Lorentz factor $W$, the density of radiated energy $E_r$, the pressure of radiated energy $P_r$ and the radiated flux $F_r$ \cite{Panchos}. The system of evolution equations has to be closed with two equations of state, for hydrodynamics and radiation. For the hydrodynamics we use the Taub-Mathews (TM) equation of state which emulates an ideal gas with adiabatic index asymptotically equal to $5/3$ in the cold regions, while in the hot regions it becomes close to $4/3$ \cite{Mignone}.  For the radiation we use the minimum entropy $M1$ closure relation because it recovers the two regimes of radiative transfer, this is, optically thick and thin \cite{Levermore,Dubroca,Gonzalez}.

These equations are solved for given initial conditions in two initially separated regions, similar to those of a shock-tube problem for the propagation of a relativistic shock, but in this case one of the regions will be stratified. Since the radiation is expected to come from regions in which the optical depth goes from high to low values \cite{Pe'er,Giannios}, the first region  corresponds to a state with constant variables and plays the role of the region where the energy of the burst comes from and is initialized  with high velocity. We use the subindex $b$ (of beam) to label the quantities in this first region. The second region corresponds to a surrounding medium which we assume to be at rest with density $\rho_m$ and pressure profiles $P_m$ that decrease with radius \cite{Mignone,DeColle}. We use the label $m$ to distinguish quantities in the surrounding medium

\begin{equation}
\rho_m=\rho_0\left(\frac{x_0}{r}\right)^{2}, \ \ P_m= P_0\left(\frac{x_0}{r}\right)^{2},
\end{equation}

\noindent where the parameters of such surrounding medium are $\rho_0=n_0m_pc^2$ with $n_0= 1\text{cm}^{-3}$, and $P_0=\rho_0c^210^{-10}$, where $m_p$ is the proton mass, $x_0$ is the location of the interface between the beam and the medium at initial time, $\rho_0$ is the density at the interface, $n_0$ is the number density of protons, $c$ the speed of light and $r$ is the radial coordinate along the propagation of the jet, that determines the density profile of the surrounding medium.

In all the simulations in this paper we assume the system to be in local thermal equilibrium and to be optically thick. Therefore the initial conditions are characterized by a set of five variables in the beam $[\rho_b,P_b,W_b,E_{r,b},F_{r,b}]$ and five on the surrounding medium $[\rho_m,P_m,W_m,E_{r,m},F_{r,m}]$. Notice that we ignore $P_r$ in both sides because we are using the M1 closure relation. We also set $W_m=0$ because the medium is assumed to be at rest initially. 

For the evolution of the system we solve the radiation relativistic hydrodynamics (RRH) system. The set of evolution equations, their relation to the above closure relations and the numerical methods used to evolve the system can be found explicitly in \cite{Panchos}.

Finally, the luminosity of the process is calculated as $L=dE_r/dt$, where $E_r$ is the radiated energy density measured by an observer located far from the interface separating the two initial states during the evolution process. We then calculate the radiated flux by diving the luminosity by the luminosity distance $d_L$  for a steady, isotropically emitting source. Thus the radiated flux is given by $F_r=L/4\pi d_L^2$ \cite{George}. This quantity defines the LC of the process.

Therefore, for each combination of initial conditions one solves the direct problem and in the end constructs a numerical LC that in the ideal case would fit the observational data of an observed LC. Nevertheless, fitting an observational LC requires to explore a parameter space with nine dimensions (recall that $W_m=0$), with each of its points  corresponding to a given parameter set of initial conditions. Each combination involves the numerical solution of a complicated system of PDEs that is time consuming. Minimizing the amount of simulations is the goal of using efficient methods to classify initial conditions, that will help at locating the appropriate region of the parameter space that is more likely to contain the parameters that fit a given LC.

\begin{figure}
\centering
\includegraphics[width= 8.9cm]{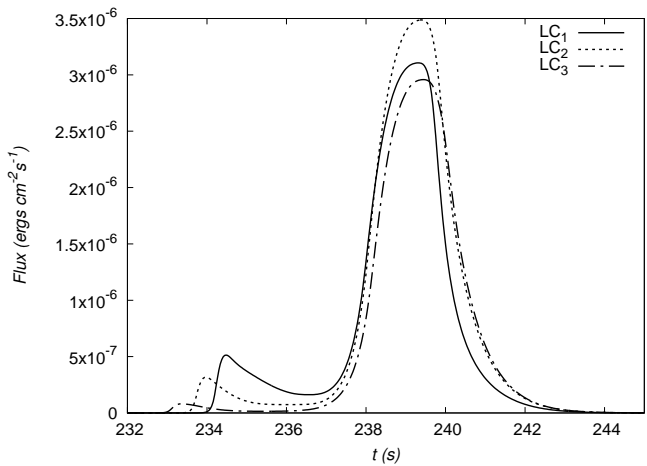}
\caption{We show a sample of signals generated with a set of parameters. The parameters used in these curves are: 
($\textrm{LC}_1$) $\rho_b =0.28\rho_0,~W_b=316.23,~E_{r,b}=1.66\times 10^{9}\text{erg/cm}^3$,
($\textrm{LC}_2$) $\rho_b =0.55\rho_0,~W_b=500,~E_{r,b}=1.66\times 10^{8}\text{erg/cm}^3$,
($\textrm{LC}_3$) $\rho_b =1.11\rho_0,~W_b=790.57,~E_{r,b}=1.66\times 10^{7}\text{erg/cm}^3$.}
\label{fig:L_sample}
\end{figure}

The problem thus would involve the construction of a classification scheme in a high dimensional space. However we have found that the most influential parameters in the LC are the density, velocity and radiation energy density of the beam  $[\rho_b,W_b,E_{r,b}]$, and this is the set of parameters we will use to illustrate our analysis. Three examples of LC are shown in Fig. \ref{fig:L_sample}, for different combinations of initial values of these physical  quantities of the injected beam. It can be seen that the signals have a prominence with a short afterglow. However they are considerably different in the sense that the amplitude and width of each of them changes when the initial parameters do. Some signals have a preglow, which highlights the time when the radiated energy density of the incident beam increases. 

It would be straightforward to find the initial conditions appropriate to fit a given observational LC, as long as it is possible to construct a catalog of LCs produced by many combinations of initial conditions of the three parameters we selected. Nevertheless, the construction of such catalog would be extremely time consuming without an educated selection of a parameter space region to start a exploration with. This is the point where a nonarbitrary and well educated location of parameters could reduce the number of simulations to be carried out. The method we propose, consists in constructing an educated set of numerical solutions of the RRH system with their respective LCs, then train, validate and test different ANNs capable to classify the physical parameters appropriate for a given observational LC.

What is expected to gain in the process is the reduction of the number of simulations required to track down the initial conditions that best fit a given LC. In the following section we will explain our implementation of the ANN method.

\subsection{Artificial neural network description}
\label{sec:nn}

In order to classify the initial data, first we fix the possible ranges of values for the three physical parameters involved in the production of the LC, i.e., $[\rho_b,W_b,E_{r,b}]$. The chosen ranges for each one of them are $\rho_b \in [0.277,1.61]\rho_0$, $W_b \in [300,900]$ and $E_{r,b} \in [8.29\times10^{9},6.71\times10^{11}]\left(\frac{M_\odot}{M}\right)^2\text{erg/cm}^3$ where $M$ is the progenitor mass associated to long GRBs. 

From each of these ranges, we select $n_p$ equidistant values of each parameter and generate $n_{p}^{3}$ numerical simulations corresponding to each combination of the possible values. The LCs obtained in the simulations are used to train the ANN. Each LC will provide data like those in Fig. \ref{fig:L_sample}, which are discretized in 220 time intervals, with the idea that the time window contains the LC data of the simulation. The value of the radiated flux is introduced as the input of the ANN. The network propagates forward the input data through a hidden layer and finally to an output layer consisting in three outputs, each one of them normalized to have a value between zero and one.

We associate each of these outputs with one among $n_c$ possible classes for each one of the parameters with $n_c \le n_p$. Basically if the first output of the ANN has a value between zero an $1/n_c$, the network is predicting that the value of $\rho_b$ used to generate that LC has a value in the first class of its range. To illustrate this, if $n_c=3$ and the three outputs of the ANN are $(0.4, 0.1, 0.8)$, this means that the corresponding classes labels of the parameters are (2,1,3). The physical values of the parameters are then within the values $\rho_b=(0.94\pm 0.22)\rho_0, W_b=400\pm100$ and $E_{r,b}=(5.61\times10^{11}\pm1.11\times10^{11})\left( M_{\odot}/M\right)^2 \text{erg/cm}^3$, where the uncertainty is computed as half of the length of the box.

To train the feedforward ANN we use an offline backpropagation algorithm with learning rate $\gamma$ that minimizes the cost function 

\begin{equation}
   C=\frac{1}{N}\sum^{N}_{p=1}\sum^3_{k=1}(T^p_k-O^p_k)^2
    \label{cost}
\end{equation}

\noindent to find the update rule for each weight during every iteration of the algorithm. Here $T^p_k$ is the $k$th target and $O^p_k$ is the $k$th output from the pattern $p$ and $N$ is the total number of LC used to train the network.

We divide the $n_{p}^{3}$ simulations in three different sets. The first one called {\it training set} contains sixty percent of the simulations and is used to adjust the network weights as explained in the previous paragraph. The second one called {\it validation set} containing twenty percent of the simulations, is used to indicate when the training process has to stop and avoid overtraining. The third and last one is called {\it prediction set}, it contains the remaining twenty percent of the simulations and is the set used to determine the quality and accuracy of the network.

\begin{figure}
\includegraphics[width=9.0cm]{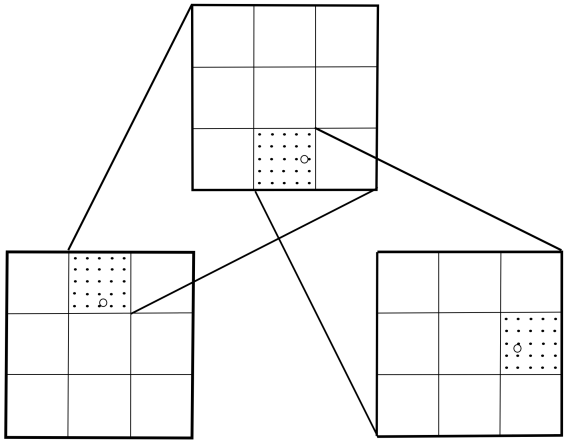}
\caption{Sampling of the parameter space with refinement showing the two dimensional parameter plane $\rho_b-W_b$. The classification is done among twenty seven cubes in reality, however we show here the two dimensional projection with $n_{c}^{2}=9$ square boxes. Each point represents a simulation with a given set of initial conditions $(\rho_b,W_{b},E_{r,b})$. In this example there are three refinement levels of the sampling. The ANN prediction is that the best parameters are in the box with coordinates $(2,3)$ in the first refinement (left-bottom), the box $(2,1)$ in the second refinement (top-center) and the box $(3,2)$ in the third refinement (right-bottom).}
\label{fig:Refinement}
\end{figure}

In order to reduce the size of the prediction's uncertainty of the parameters, once the network has selected a given interval for the three physical parameters, we perform another set of simulations contained inside such selected box. This means we produce another set of $n_{p}^{3}$ simulations within the selected box of initial conditions and repeat the process of classification, but notice that this time within a smaller box in the parameter space. This refinement process is illustrated in Fig. \ref{fig:Refinement} for three refinement levels. Every time we repeat this process the uncertainty of the initial conditions decreases. In the results presented in the following section, we apply this process three times.
 
\section{Results}
\label{sec:results}

In order to find out an optimal performance of our ANN, we explore different topologies, by changing the learning rate and the number of hidden neurons. We trained the network during $2\times10^4$ iterations for all the different configurations. We have chosen two different learning rate constants ($\gamma=1\times10^{-3}$ and $\gamma=2\times10^{-3}$). We have combined each one of them with 5, 10, 20 and 40 hidden neurons, one at a time. We carried out two experiments to test the efficiency of each of above-mentioned topologies. In the first one, we divided the ranges of rest mass density,  Lorentz factor and radiated energy density, in $n_p=12$ equidistant values of each parameter, creating a total of $12^3=1728$ LCs. With these 12 values, we analyzed 3, 6 and 12 classes for each parameter.  In Table \ref{table:classes} we show the percentages of correct classifications for each scenario with two different values of the learning rate constant.

We can see that if the number of hidden neurons increases, the accuracy of the prediction increases too, but also the computational cost as seen in Table \ref{table:ctime}. This behavior happens in general for all the scenarios independently of the number of classes. For the same network parameters, if we use a smaller number of classes, the accuracy of the prediction increases but also the uncertainty of the estimated parameters related with the size of the interval used to classify.

In the second experiment, we divided the ranges of physical parameters, into $n_p=16$ equidistant values of each parameter, generating a total of $16^3=4096$ LCs. With these 16 values we form 4, 8 and 16 classes for each parameter. The predictions for this experiment are shown in Table \ref{table:classes16}. Like in the previous experiment, the accuracy of the network increases as we increase the number of neurons. It is worth noticing that this improvement in accuracy reaches, in both experiments, a maximum with a number of neurons  between 20 and 40. We also show the computing time required for the classification in Table \ref{table:ctime}.

Comparing these two experiments, we can see that predictions in the second experiment are less precise than in the first one. Also in Tables \ref{table:classes} and \ref{table:classes16} we can notice that the number of correct classifications is bigger for experiment one than it is for experiment two, nevertheless in the first (second) one the size of the intervals used to classify is bigger (smaller) which implies the parameter's uncertainties are bigger (smaller).

After the network is trained, we evaluate its performance using the prediction set. The data corresponding to this set is propagated through the network using the trained weights and we found that the network classifies correctly the data for each output at least 84.8\% percent of the times, as can be seen in the first row of Table \ref{table:nstructure6}. In order to quantify the quality of our predictions, we also prepared two tests based on numerical and observational  LCs.

As mentioned in the previous section, to reduce the uncertainty associated to parameters estimates, we perform a second classification assuming a smaller box of physical parameter values centered in the predicted class. Then a new set of $n_{p}^{3}$ LCs are prepared for the ANN. The results are presented in Tables \ref{table:nstructure} and \ref{table:nstructure6} for a different number of hidden neurons and learning constant values.

\begin{table}[]
\begin{center}
\begin{tabular}{|c|c|c c c|c c c|c c c|}
\cline{3-11}
\multicolumn{2}{c}{ }&\multicolumn{9}{c}{Classes} \\ \cline{3-11}
 \multicolumn{2}{c|}{ }& \multicolumn{3}{|c|}{3}& \multicolumn{3}{|c|}{6}& \multicolumn{3}{|c|}{12} \\ 
 \hline
 $\gamma$&H&$\rho_b$&$W_{b}$&$E_{r,b}$&$\rho_b$&$W_{b}$&$E_{r,b}$&$\rho_b$&$W_{b}$&$E_{r,b}$  \\ \hline
 &5& 80.6 &77.1&96.2&63.4&62.6&85.7&  55.3 & 35.3 & 60.2 \\
 .001&10&98.2&84.3&96.2&72.4&66.9&89.2& 78.8  &45.2  & 68.9  \\  
 &20&97.3&88.9&96.8&71.8&79.1&95.9&74.2& 55.3 &  79.4  \\
  &40& 91.8 & 88.6 & 97.3  & 76.8 & 77.6 & 97.3 & 82.0 & 50.1 & 86.3\\ \hline
  &5& 84.6 & 86.3 &95.6  & 63.1 &52.4  & 78.2  & 54.2  & 32.7 & 61.1  \\
 .002&10&98.2 &84.0  &95.0  & 78.8  & 70.7 & 91.0  & 73.6  & 49.5 & 75.6  \\ 
 &20& 96.2 & 86.9 & 96.2 &  81.4 & 78.8 & 96.5  & 85.5  & 53.9 & 88.6\\
  &40& 97.6 & 87.5 & 99.4 & 80.2 & 84.0 & 97.9 & 90.1 & 55.9 & 87.8\\ \hline
\end{tabular}
\caption{Percentage of correct classifications for each physical parameter for two values of the learning rate $\gamma$, the number of hidden neurons $H$ and the number of classes for experiment one,  where the physical parameters were divided in $12$ equidistant values.}
\label{table:classes}
\end{center}
\end{table}

\begin{table}[]
\begin{center}
\begin{tabular}{|c|c|c c c|c c c|c c c|}
\cline{3-11}
\multicolumn{2}{c}{ }&\multicolumn{9}{c}{Classes} \\ \cline{3-11}
 \multicolumn{2}{c|}{ }& \multicolumn{3}{|c|}{4}& \multicolumn{3}{|c|}{8}& \multicolumn{3}{|c|}{16} \\ 
 \hline
 $\gamma$&H&$\rho_b$&$W_{b}$&$E_{r,b}$&$\rho_b$&$W_{b}$&$E_{r,b}$&$\rho_b$&$W_{b}$&$E_{r,b}$  \\ \hline
 &5&79.8 & 81.7 &  93.5 & 56.8 & 58.6 & 83.4 & 49.0 & 30.9 & 65.2 \\
 .001&10& 82.1 & 86.4 & 98.0 & 61.4 & 62.1 & 84.0 & 68.1 & 40.7 & 73.6\\  
 &20& 85.8 & 86.3 & 96.0 & 69.1 & 69.2 & 88.2 & 67.5 & 46.2 & 74.0 \\
   &40& 85.6 & 86.9 & 97.6 & 71.9 & 70.8 & 91.0 & 76.3 & 52.5 & 83.1\\ \hline
  &5& 81.3 & 84.5 & 94.6 & 70.6 & 58.7 & 73.7 & 56.4 & 34.5 & 51.7  \\
 .002&10& 85.1 & 85.3 & 95.1 & 65.8 & 64.1 & 79.5 & 67.0 & 36.9 & 77.6 \\ 
 &20& 85.7 & 87.4 & 98.9 & 76.2 & 68.2 & 91.8 & 79.2 & 46.3 & 77.9\\
   &40& 81.0 & 87.6 & 97.9 & 72.5 & 64.1 & 89.6 & 78.6 & 49.5 & 83.5\\ \hline
\end{tabular}
\caption{Percentage of correct classifications for each physical parameter for two values of the learning rate constant $\gamma$, the number of hidden neurons $H$ and the number of classes in the second experiment, where the physical parameters were divided in $16$ equidistant values.}
\label{table:classes16}
\end{center}
\end{table}

\begin{table}[]
\begin{center}
\begin{tabular}{|c|c | c|}
\hline
&\multicolumn{2}{|c|}{Computational time (s)}\\ \hline
H & $n_p=12$ & $n_p=16$ \\ \hline
5 & 8167 & 19540 \\
10 & 16400 & 40117 \\
20 & 32395 & 81946 \\
40 & 74282 & 174554 \\
\hline
\end{tabular}
\caption{Computational time measured in seconds for each network structure, when $n_p= 12$ and 16. }
\label{table:ctime}
\end{center}
\end{table}

\begin{table}[]
\begin{center}
\begin{tabular}{|c| c c c |c|}
\hline
H &$\rho_b$&$W_{b}$&$E_{r,b}$& Computational\\
 &\%&\%&\% & time (s)\\
\hline \hline
1 & 36.5 & 33.7 & 37.2 & 700\\ \hline
2 & 48.9 & 33.7 & 75.1 & 1411\\ \hline
3 & 88.9 & 33.7 & 98.6 & 2189\\ \hline
4 & 95.1 & 64.1 & 99.3 & 2843\\ \hline
5 & 97.2 & 72.4 & 98.6 & 3536\\ \hline
6 & 97.9 & 70.0 & 100 & 4384\\ \hline
7 & 95.1 & 73.1 & 97.9 &4877\\ \hline
8 & 97.9 & 76.5 & 100 & 5802\\ \hline
9 & 97.9 & 73.1 & 100 & 6513\\ \hline
10 & 96.5 & 78.6 & 100 & 6899\\ \hline
20 & 98.6 & 80.6 & 100 & 14169\\ \hline
40 & 97.2 & 82.7 & 100 & 28325\\ \hline
60 & 97.9 & 84.8 & 100 & 42971\\ \hline
80 & 97.2 & 84.1 & 100 & 56713\\ \hline
100 & 97.9 & 84.8 & 100 & 70324\\ \hline
\end{tabular}
\caption{Percentage of correct classifications for each physical parameter using the learning rate $\gamma=0.001$ and varying the number of hidden neurons $H$ for $n_c=3$ and $n_p=9$, including the computational time measured in seconds, used by the network for the first refinement.}
\label{table:nstructure}s
\end{center}
\end{table}

\begin{table}[]
\begin{center}
\begin{tabular}{|c| c| c c c |}
\hline
$\gamma$&H &$\rho_b$&$W_{b}$&$E_{r,b}$\\
\hline \hline
&5 & 95.1 & 84.8 & 98.6 \\ 
.0025&20 & 98.6 & 85.5 & 100\\ 
&60 & 97.9 & 91.7 & 100 \\ \hline
&5 & 97.2 & 72.4 & 98.6 \\ 
.00075&20 & 96.5 & 77.2 & 99.3\\ 
&60 & 97.2 & 82.0 & 100 \\ \hline
\end{tabular}
\caption{Percentage of correct classifications for each physical parameter varying the learning rate $\gamma$ and considering $H$ hidden neurons for $n_c=3$ and $n_p=9$. }
\label{table:nstructure6}
\end{center}
\end{table}

\subsection{Results with numerically generated LCs}

We produce numerically two LCs with the only condition that the parameters to generate such LCs lie within the region containing the parameters used to train the network. The specific data used are shown in Table \ref{table:tets}. In both tests we set the progenitor mass to $10M_\odot$. 

\begin{table}[htbp]
\begin{center}
\begin{tabular}{|c|c|c|c|}
\hline
Test & $\rho_b \ (\text{g}/\text{cm}^3)$ & $W_b$ & $E_{r,b} \ \left(\frac{M_\odot}{M}\right)^2(\text{erg/cm}^3)$ \\
\hline \hline
1 & $1.54\rho_0$ & $790.5$ & $6.63\times10^{11}$\\ \hline
2 & $1.58\rho_0$ & $301.5$ & $1.66\times10^{11}$\\ \hline
\end{tabular}
\caption{Initial parameters for two particular initial conditions whose LCs will be used to determine the accuracy of the classification scheme.}
\label{table:tets}
\end{center}
\end{table}

We remind the reader that the parameter box has parameter values $(\rho_p,W_b,E_{r,b})\in[0.277,1.61]\rho_0\times[300,900]\times[8,29\times 10^{9},6.71\times 10^{11}](M_{\odot}/M)^2$. 
Assuming $n_p=9$ and $n_c=3$, the initial conditions for case 1 in Table \ref{table:tets} would belong to the box (3,3,3) within the first refinement, and the boxes (3,2,3)-(2,3,3) in the second and third refinements. Therefore the perfect classifier should find that the LC was generated with parameters within the box sequence (3,3,3)-(3,2,3)-(2,3,3) for Test 1. On the other hand, for Test 2 the sequence should be (3,1,1)-(3,1,3)-(3,1,1).

The ANN predicted the following box sequences (3,3,3)-(2,1,3)-(1,3,3) and (3,1,1)-(3,1,3)-(3,1,1) for  Tests 1 and 2 respectively.  We present in Table \ref{table:class}, the explicit parameter values and their uncertainty, defined as half of the length of the class in the third refinement.

\begin{table}[htbp]
\begin{center}
\begin{tabular}{|c|c|c|c|}
\hline
Test & $\rho_b \ (\text{g}/\text{cm}^3)$ & $W_b$ & $E_{r,b} \ \left(\frac{M_\odot}{M}\right)^2(\text{erg/cm}^3)$ \\
\hline \hline
1 & $1.34\pm0.024 \rho_0$ & $755.55\pm11.11$ & $6.59\pm 0.123\times10^{11} $\\ \hline
2 & $1.59\pm0.024 \rho_0$ & $311.11\pm11.11 $ & $1.68\pm 0.123\times10^{11} $\\ \hline
\end{tabular}
\caption{Predictions obtained with the ANN after three levels of refinement for each test.}
\label{table:class}
\end{center}
\end{table}

\begin{figure}
	\centering
\includegraphics[width=8.9cm]{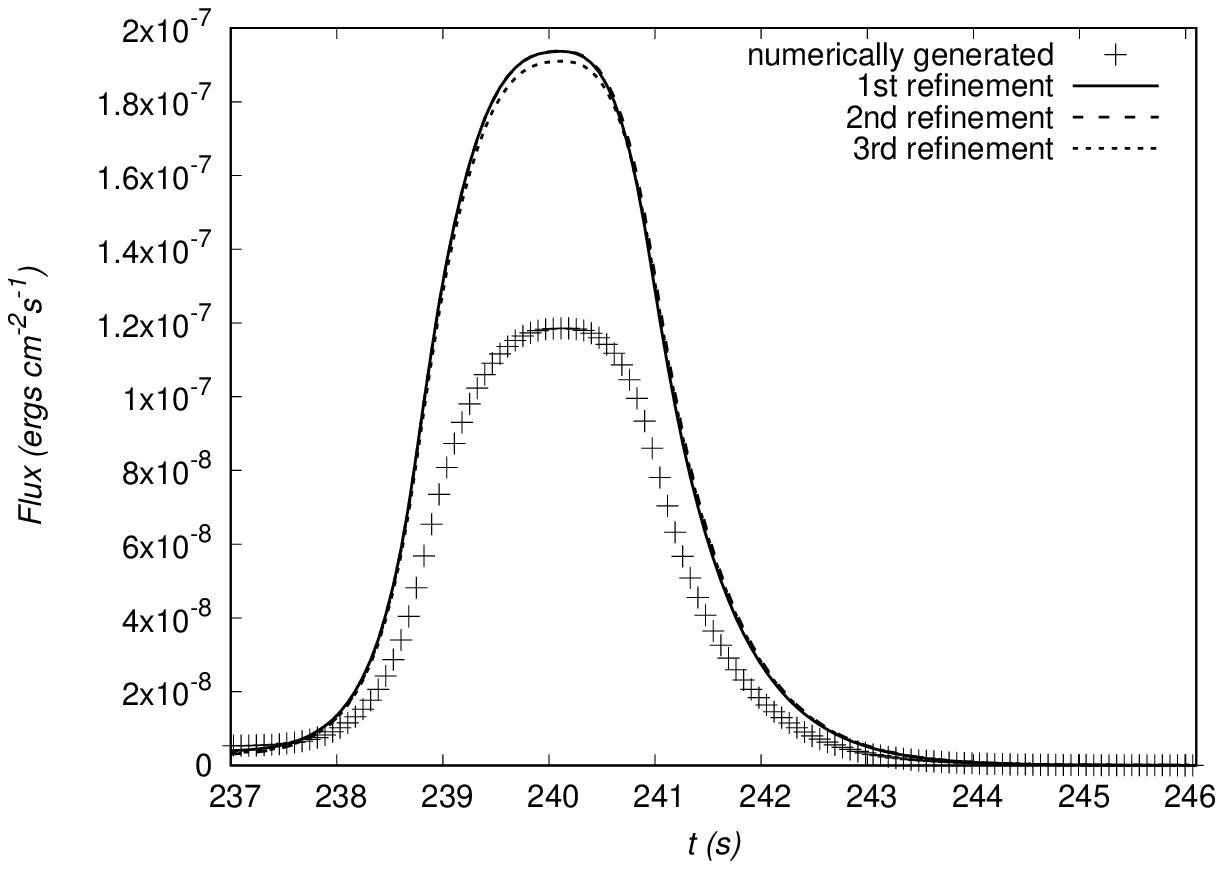}
\includegraphics[width=8.9cm]{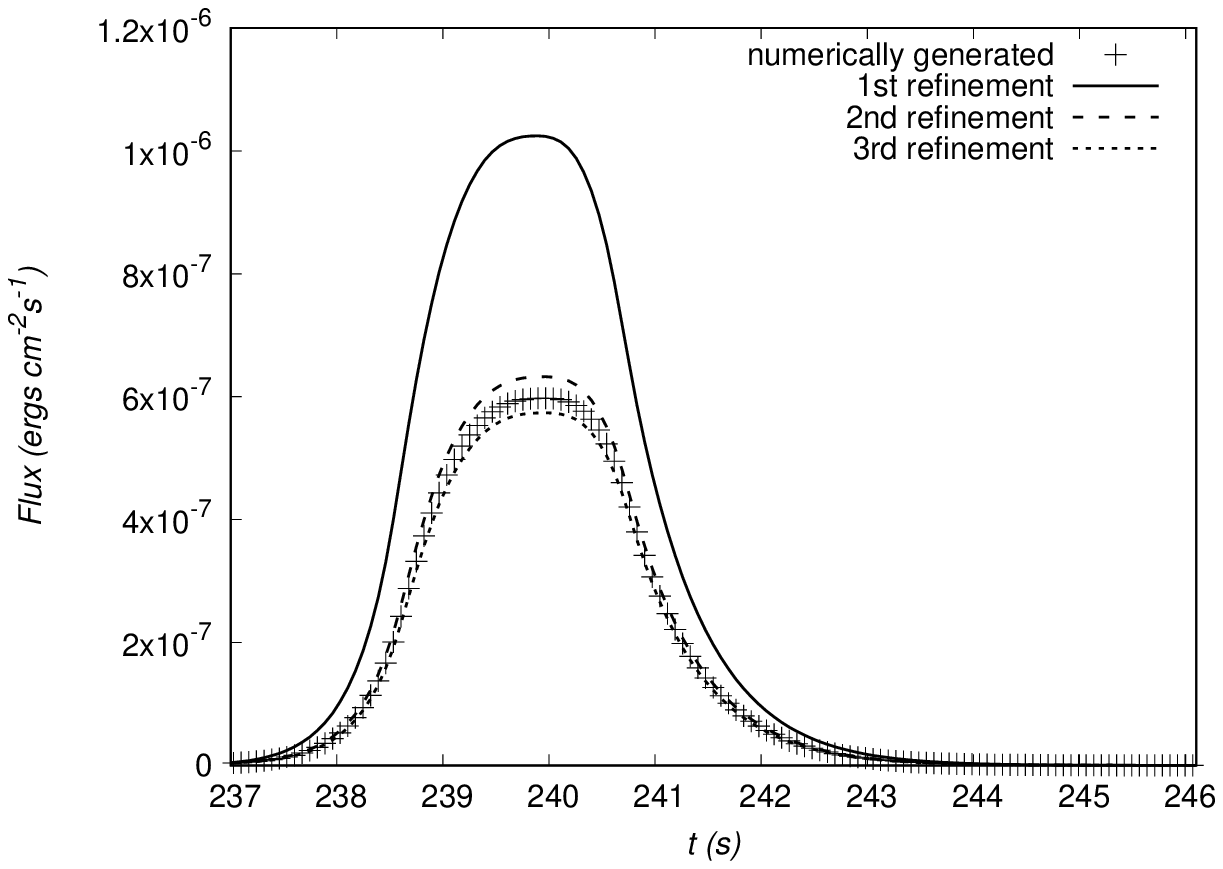}
\caption{\label{fig:test} The LC generated numerically for Test 1 (top) and Test 2 (bottom), compared with the predictions of the ANN. The first case illustrates one of the scenarios where the ANN fails at classifying the initial conditions in the second refinement. The second test illustrates a successful case.}
\end{figure}

For the sake of illustration we have chosen Test 1 as an example showing the ANN failing at classifying in the second refinement, notice that the density fails by a considerable 15\%. Test 2 is an example of a successful classification where the density shows an error less than 1\%.  In Fig. \ref{fig:test} we compare the original LC and the one predicted by classification of parameters by the ANN of these two tests.

\subsection{Results with observed GRBs}

\begin{figure*}
	\centering
\includegraphics[width=8.9cm]{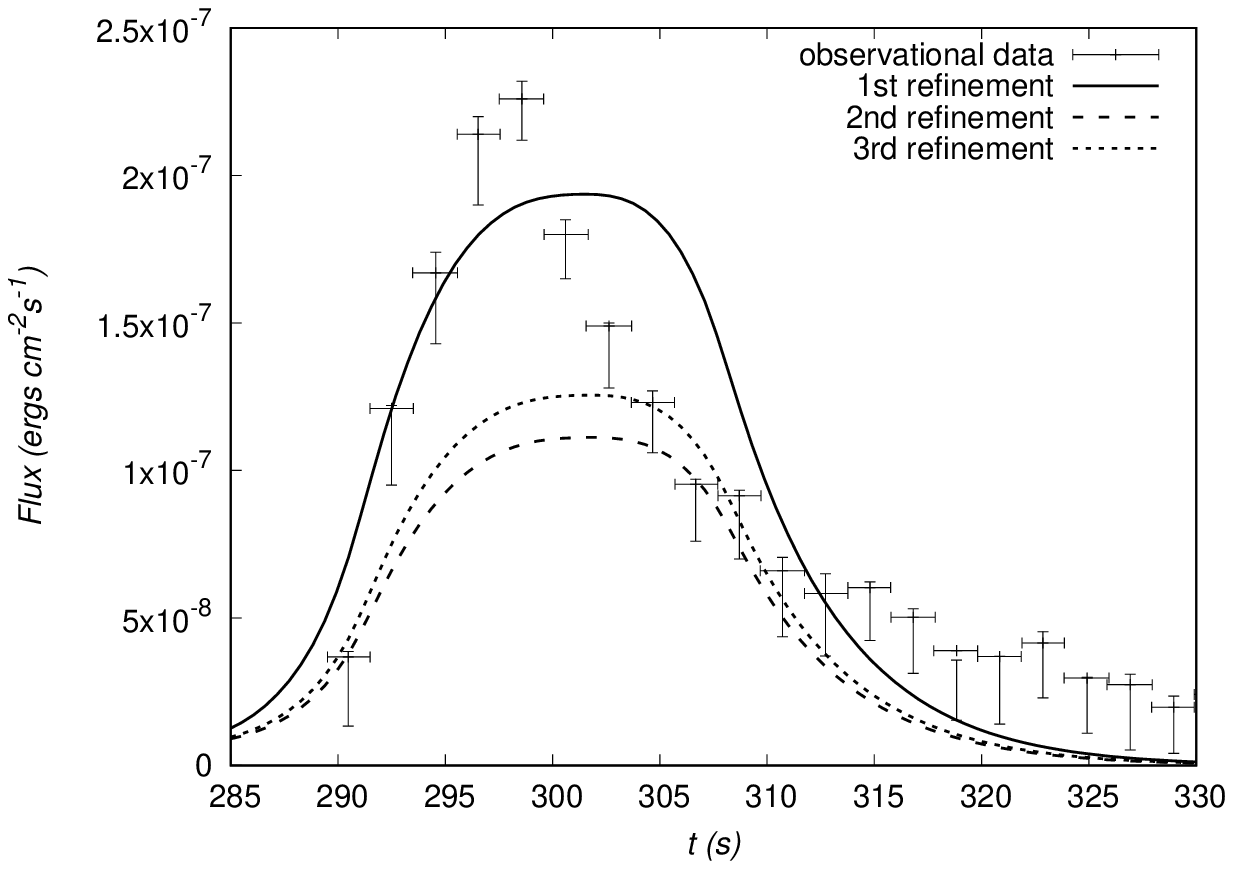}
\includegraphics[width=8.9cm]{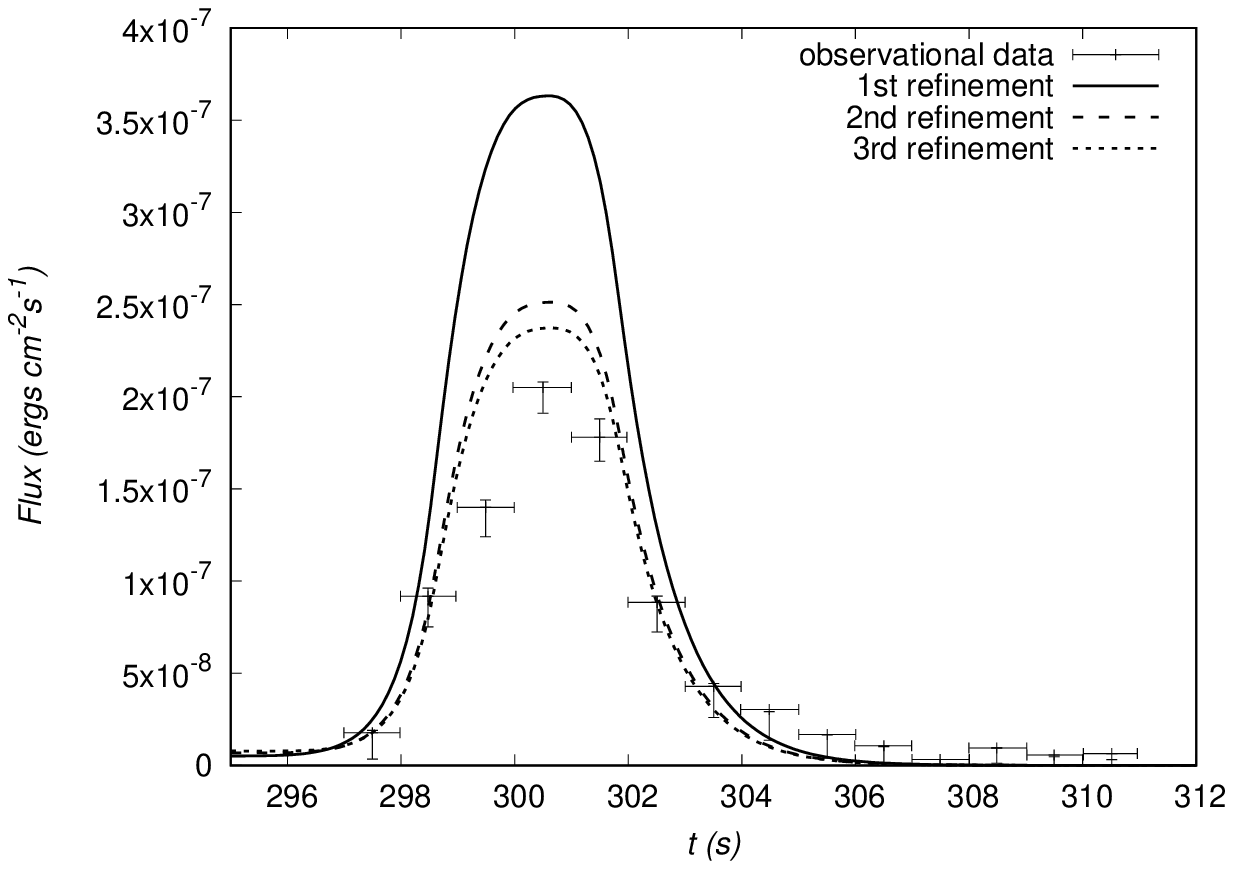}
\includegraphics[width=8.9cm]{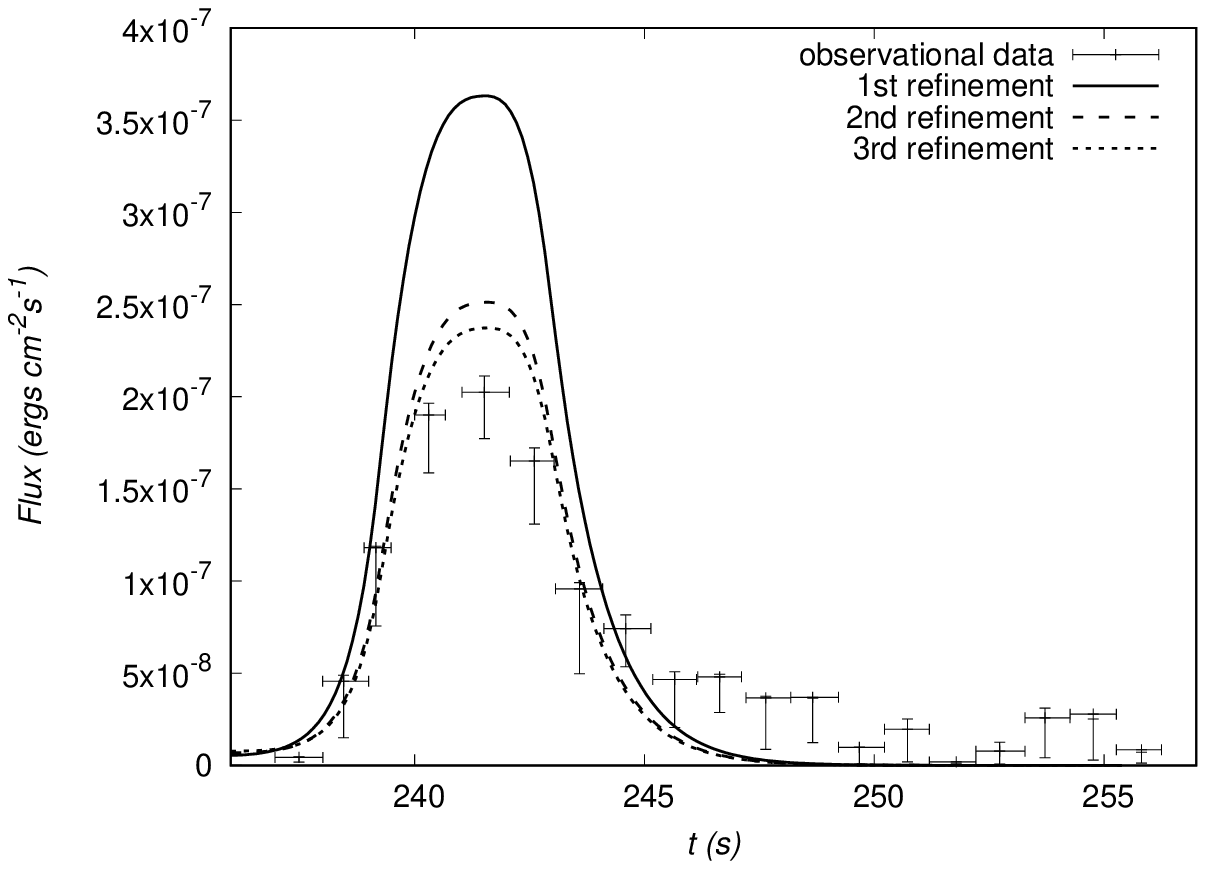}
\includegraphics[width=8.9cm]{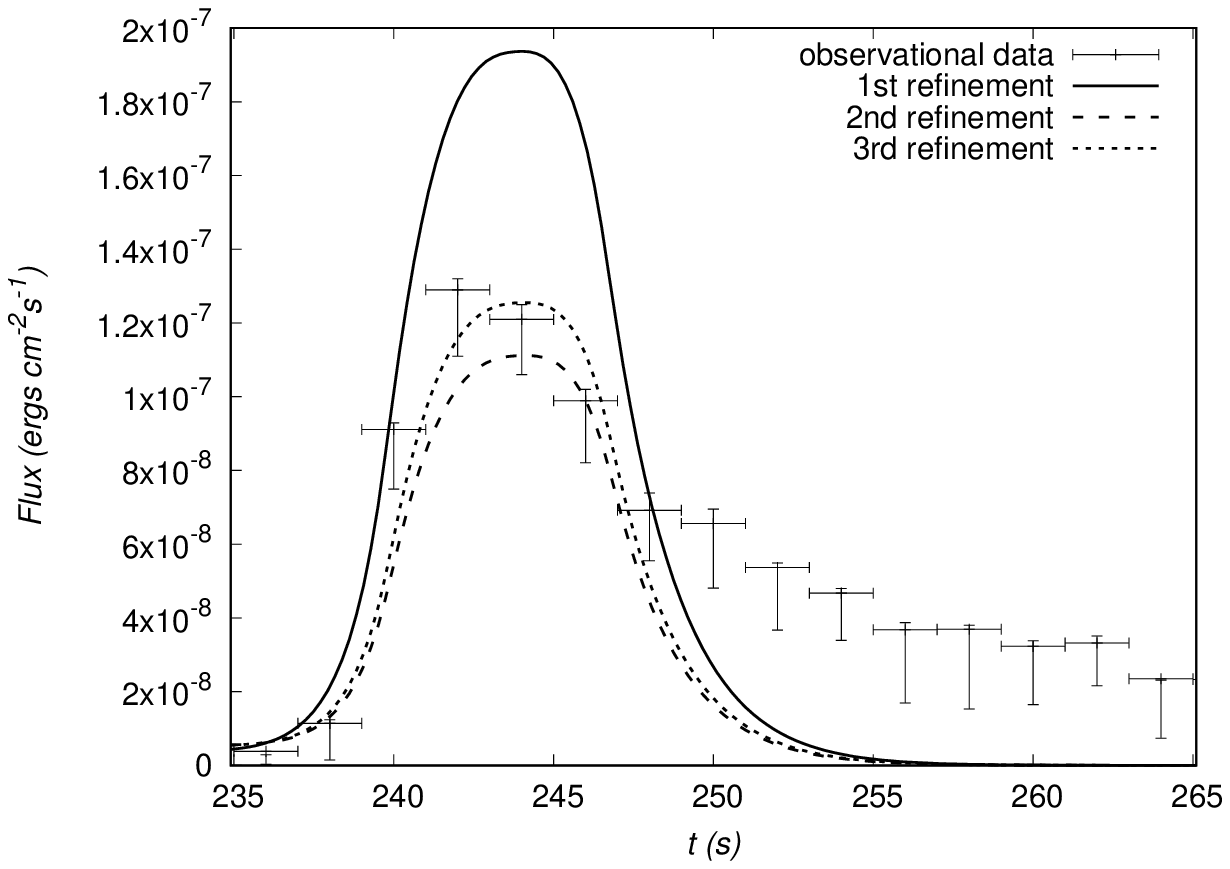}
\includegraphics[width=8.9cm]{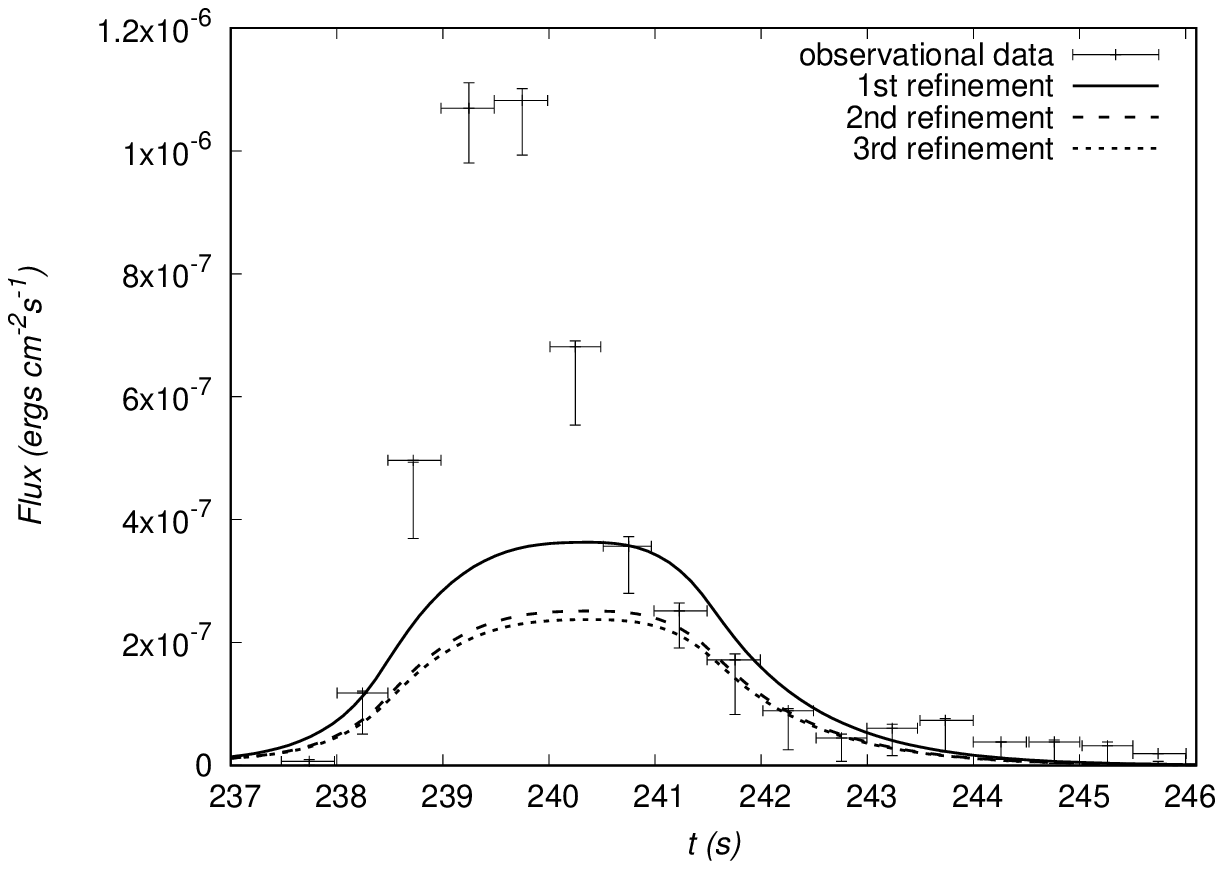}
\caption{\label{fig:GRBs} The light curve data of five long GRBs, from left to right and top to bottom: GRB051111, GRB060206, GRB060904B, GRB070318, and GRB080413B respectively. The data represented by dots with error bars were taken from \cite{Mendoza}. The three lines of each plot correspond to the predictions in the first, second and third refinement.}
\end{figure*}

Before classifying the parameters we prepare LC data as a time-series that will feed our ANN. For this we take observational data, like those in Fig. \ref{fig:GRBs} and convert them into code units. We fit the data with a high order polynomial that we use to generate a time series in a discrete time domain with uniform resolution $\Delta t$ used for the light curves we generated numerically in the previous subsection. When this process is done, we introduce the data into the trained ANN and proceed to the classification.

As we initialize the ANN weights with random numbers, the network's prediction may change when the initial weights are different, this implies that training depends on the initial weights. In order to overcome the dependence on this randomness, we train ten networks with different initial weights. To know what prediction was the best, the ten networks were trained and the coordinates that were predicted more times and those that best fit in a given refinement were selected for the next refinement. We used three  refinements and the predictions for each of the observed light curves are shown in Table \ref{initial_condition}. The GRBs are: GRB051111, GRB060206, GRB060904B,GRB070318 and GRB080413B. These long GRBs were taken on an energy band $15-150$ KeV from \cite{Mendoza}. 

In Fig. \ref{fig:GRBs} we show the fits for these five long GRBs. The parameters $\rho_b,~W_b,~E_{r,b}$, used to produce these plots are those of the center of the box selected by the ANN. In the first case, we show how in the first refinement the fit is better than in the subsequent refinements. This is a method to lock down further refinements. The following three curves show a reasonable fit with the third refinement. Finally the fit of the fifth curve is a failure in the classification.

\begin{table}
\begin{center}
\begin{tabular}{c  c  c  c  c}
\hline
\hline
\hline
Refinement &  $\rho_b(\text{g}/\text{cm}^3)$  & $W_b$  & $E_{r,b}(\text{erg}/\text{cm}^3)$  & $M(M_\odot)$ \\
\multicolumn{5}{l}{GRB051111} \\
\hline
 $(3,3,3)$         & $1.39\pm0.22$           &$800\pm100$       & $5.61\pm1.11$  & $25$   \\
\hline
 $(1,3,3)$         & $1.24\pm0.073$          &$866.6\pm33.33$   & $6.34\pm0.368$  & $25$   \\
\hline
 $(1,2,3)$         &$1.19\pm0.024$           &$866.6\pm11.11$   & $6.59\pm0.123$  & $25$  \\
\hline
\multicolumn{5}{l}{GRB060206} \\
\hline
 $(3,1,3)$         &  $ 1.39\pm0.22$         &$400\pm100$       & $5.61\pm1.11$  & $12.5$\\
\hline
 $(3,3,1)$         &  $ 1.54\pm0.073$        &$466.6\pm33.33$   & $4.87\pm0.368$  & $12.5$ \\
\hline
 $(3,1,3)$         &  $ 1.59\pm0.024$         &$444.4\pm11.11$   & $5.11\pm0.123$  & $12.5$ \\ 
 \hline
 \multicolumn{5}{l}{GRB060904B} \\
 \hline
 $(3,1,3)$         &  $ 1.39\pm0.22$         &$400\pm100$       &  $5.61\pm1.11$ & $15$ \\
 \hline
 $(3,3,1)$         & $ 1.54\pm0.073$         &$466.6\pm33.33$   & $4.87\pm0.368$  & $15$  \\
 \hline
 $(3,1,3)$        &  $ 1.59\pm0.024$          &$444.4\pm11.11$   &  $5.11\pm0.123$ & $15$  \\
 \hline
 \multicolumn{5}{l}{GRB070318} \\
 \hline
  $(3,3,3)$        &  $1.39\pm0.22$          &$800\pm100$       & $5.61\pm1.11$  & $20$ \\
 \hline
  $(1,3,3)$        &  $1.24\pm0.073$         &$866.6\pm33.33$   & $6.34\pm0.368$  & $20$  \\
 \hline
  $(1,3,3)$        &  $1.19\pm0.024$         &$888.8\pm11.11$   & $6.59\pm0.123$  & $20$  \\
 \hline
 \multicolumn{5}{l}{GRB080413B} \\
 \hline
  $(3,1,1)$         & $1.39\pm0.22$         &$400\pm100$       &  $3.4\pm1.11$  & $10$ \\
 \hline
  $(3,1,3)$         & $1.54\pm0.073$        &$333.3\pm33.33$   &  $1.92\pm0.368$  & $10$  \\
 \hline
  $(1,2,1)$         & $1.49\pm0.024$        &$333.3\pm11.11$   & $1.68\pm0.123$   & $10$  \\
\hline
\hline
\end{tabular}
\caption{\label{initial_condition} Parameters of the initial jet. In order for the values of the rest-mass density and radiative energy density to be correctly rescaled, they must be multiplied by $\rho_0$ and $\left(\frac{M_\odot}{M}\right)^2\times10^{11}$ respectively.}
\end{center}
\end{table}

\section{Final comments}
\label{sec:final}

We have presented a tool that classifies the light curves generated by a single one-dimensional relativistic jet produced by different values of the rest mass density, Lorentz factor and radiated energy of the beam, based on a radiation relativistic hydrodynamical model.

As shown in \cite{Panchos} the model used to generate the jets, is still a simple one, however useful at fitting the amplitude of LGRB flux light curves. Even though the model does not contain all the ingredients of the most sophisticated case, it helps at illustrating the usefulness of the ANNs as a classifier of initial conditions in this specific astrophysical scenario.

We tested our method using numerically generated LCs and obtained an 84.8\% of accuracy in the classification as shown in Table \ref{table:nstructure6}. We show the Test 1 as a case where the ANN fails at classifying properly. We also applied the method to observational LCs, where the method was correct in four out of five cases analyzed. Finally, because the uncertainty on the parameters depends on the refinement of the parameter space, it is expected that by increasing the number of refinements, the size of the quality of the fit will improve.

Using a single opacity associated with a single emission mechanism, in our case  bremsstrahlung radiation and Thomson scattering cannot explain the two regimes of the LC. A more realistic scenario  would include the inverse Compton process acting on a higher energy regime and the Thompson channel on the low energy regime. This can be overtaken by using a multiband or multifrequency model that can take into account two or more processes with the same code (e.g. \cite{valencia}), a case that we will implement in the future.

In conclusion, our method provides a straightforward way to track down the parameters within a given accuracy. Similar astrophysical problems could require the solution of three dimensional evolution equations, which require a considerable computer power and therefore the exploration of the parameter space would be very costly. In such scenario, our approach could show its strength by using a systematic exploration which is expected to be more efficient than using brutal force and thus a costly exploration. We are aware of the limitations of our method, some of them include the need of producing a considerable set of simulations that eventually could be computationally prohibitive, and the possible failures in zones of the parameter space where the problem might seem degenerate to the network. Nevertheless, these kind of approaches are essential at solving inverse problems involving models related to Partial Differential Equations.


\section*{Acknowledgments}
This research is supported by grants CIC-UMSNH-4.9, CIC-UMSNH-4.23  and CONACyT 258726 (Fondo Sectorial de Investigaci\'on para la Educaci\'on). Part of the simulations were carried out in the computer farm funded by CONACyT 106466. The authors also acknowledge the computer resources, technical expertise and support provided by the Laboratorio Nacional de Superc\'omputo del Sureste de M\'exico, CONACyT network of national laboratories. We also thank ABACUS Laboratorio de Matem\'aticas Aplicadas y C\'omputo de Alto Rendimiento del CINVESTAV-IPN, grant CONACT-EDOMEX-2011-C01-165873, for providing computer resources.


\end{document}